\documentclass{elsart}

\usepackage{natbib}

 \usepackage{epsfig}

\usepackage{amssymb}

\newcommand{\chan}{{\sl Chandra}}

\newcommand{\farcs}{\hbox{$.\!\!^{\prime \prime}$}}
\newcommand{\farcm}{\hbox{$.\!\!^{\prime}$}}
\def\aciss3{{ACIS-S3}}

\begin{document}

\begin{frontmatter}



\title{The jets of the Vela pulsar.}


\author{O.\ Y.\ Kargaltsev, G.\ G.\ Pavlov, M.\ A.\ Teter, and D.\ Sanwal}

\address{The Pennsylvania State University, 525 Davey Lab,
University Park, PA 16802, USA}

\begin{abstract}

{\sl Chandra} observations of the Vela pulsar-wind nebula (PWN) have
revealed a jet in the direction of the pulsar's proper motion, and
a counter-jet in the opposite direction, embedded in diffuse
nebular emission. The jet consists of a bright, $8''$-long inner
jet, between the pulsar and the outer arc, and a dim, curved outer
jet that extends up to $\sim 100''$ in approximately the same
direction.
 From the analysis of thirteen {\sl Chandra} observations
spread over $\approx 2.5$ years we found that this outer jet shows
particularly strong variability, changing its shape and
brightness. We observed bright blobs in the outer jet moving away
from the pulsar with apparent speeds (0.3--$0.6)\, c$ and fading
on time-scales of days to weeks. The spectrum of the outer jet
fits a power-law model with a photon index $\Gamma=1.3\pm0.1$. For
a distance of 300 pc, the apparent average luminosity of the outer
jet in the 1--8 keV band is about $3\times 10^{30}$ erg s$^{-1}$,
compared to  $6\times 10^{32}$ from the whole PWN within $42''$
from the pulsar. The X-ray emission of the outer jet can be
interpreted as synchrotron radiation of ultrarelativistic
electrons/positrons. This interpretation allows one to estimate
the magnetic field, $\sim 100$ $\mu$G, maximum energy of X-ray
emitting electrons, $\sim 2\times 10^{14}$ eV, and energy
injection rate, $\sim 8\times 10^{33}\, {\rm erg\, s}^{-1}$, for
the outer jet. In the summed PWN image we see a dim,
$\sim2'$-long outer counter-jet, which also shows a power-law
spectrum with $\Gamma \approx 1.2$--1.5. Southwest of the
jet/counter-jet (i.e., approximately perpendicular to the
direction of pulsar's proper motion), an extended region of
diffuse emission is seen. Relativistic particles responsible for
this radiation are apparently supplied by the outer jet.

\end{abstract}


\end{frontmatter}

\section{Introduction}
\label{}

Recent detailed images obtained with {\sl Chandra} for PWNe around
the Crab pulsar (Hester et al.\ 2002), Vela pulsar (Helfand et al.
2001; Pavlov et al.\ 2001), and PSR B1509--58 (Gaensler et al.\
2002) have shown approximately axially-symmetric PWN morphologies,
with an extended jet-like structures stretched along the symmetry
axis. This suggests that the jets are common to at least young
pulsars and are not limited to accreting systems (e.g., AGNs,
microquasars). Most likely, pulsar jets are associated with
collimated outflows of relativistic particles along the pulsar's
rotation axes.

Due to its proximity ($d\simeq 300$ pc; Caraveo et al. 2001), the
Vela PWN is particularly well suited for studying the pulsar
outflows. We carried out a series of eight monitoring observations
with the {\sl Chandra} Advanced CCD Imaging Spectrometer (ACIS).
 These observations
 have confirmed the dynamical
structure of the PWN (Pavlov et al.\ 2001), with most dramatic
changes occurring in the outer jet. Moreover, we were able to
detect an ``outer counter-jet'', a much dimmer extension of the
southeast (counter-)jet.  Here, we focus on the highly variable
outer jet, which has been detected in ten ACIS observations
(carried out from 2000 April 30 through 2002 August 6) and three
observations with the High Resolution Camera (HRC).



\section{Results}

The large-scale X-ray structure of the Vela PWN is shown in the
summed image composed of the last eight ACIS observations (upper left
 panel of Fig.\
1; see Pavlov et al.\ 2002 for technical details of the observations,
 image reduction
and co-alignment). This deep image clearly
reveals a long
outer jet [7],
approximately in the direction of the proper motion (PA $\simeq
307^\circ$; Caraveo et al. 2001). The outer jet extends for
$\approx 1\farcm 7 = 0.14\, d_{300}$ pc away from the pulsar, where $d_{300}$
is the distance to the pulsar in units of 300 pc. The
characteristic width (diameter) of the outer jet is about $3
\times 10^{16}d_{300}$ cm. A much fainter outer counter-jet [8] is
seen in the opposite direction. Also, we see
extended diffuse emission [9] southwest of the jet/counter-jet
line, which is obviously connected to the ``main body'' of the
PWN. The inner PWN with the pulsar [1] at its center consists of
the inner arc [2], the outer arc [3], the inner jet [4], and the
inner counter-jet [5]. The inner jet is directed northwest from
the pulsar in the direction of the
pulsar's proper motion, and the counter-jet is directed toward the
southeast (PA $\simeq 127^\circ$). The bright PWN core (white in Fig.\ 1) is
 surrounded by a ``shell'' [6] of diffuse
emission. The outer jet looks like an extension
of the much brighter inner jet, well beyond the apparent
termination point of the inner jet at its intersection with the
outer arc. Finally, Figure 1 demonstrates that the outer jet
preferentially bends to the south-west of the jet/counter-jet line
and apparently connects to the extended diffuse emission region
(Fig.\ 1; upper left panel).

Figure 1 also
presents the time sequence of the \aciss3 and HRC-I observations
for the outer jet. The variability of the outer jet is clearly
seen, in both the HRC and ACIS images. Over the thirteen
observations, with different periods of time between each of them,
we distinguish three different types of variability. First, the outer
 jet shifts from side to side, bending and
apparently twisting.  Second, the blobs move outward along the
outer jet. Finally, the blobs change in brightness and eventually
disappear.

The most dramatic variations we see are the large-scale
bends of the jet (e.g. compare observations 1, 2, 3
and 13 in Fig.\ 1). The short time (16 days) between the fifth and
eighth observations suggests that this bending
occurs on
 a time
scale of order of weeks. The
  small-scale changes are seen in all of the
observations.  For instance, the ``base'' of the outer jet (where
it leaves the shell
--- see the white boxes in Fig.\ 1 in upper right and bottom panels)
shifts from one observation to the next. Typical apparent speeds
of these shifts are of order a few tenths of speed of light. The
apparent speeds of blobs A and B are $(0.35\pm 0.06)\, c$ and
$(0.51\pm 0.16)\, c$, respectively. Blob C vanished  quickly,
having an (unconstrained) apparent speed of $(0.6\pm 0.7)\,c$.
Thus, the observed variations suggest typical bulk flow velocities of 0.3--0.7 of
the speed of light.

 The outer jet is, on average, a factor of 7 brighter than
the outer counter-jet. If the outer jet and outer counter-jet are
intrinsically similar but streaming along a straight line (on average) in
 opposite directions, then
the difference in brightness means that the outer jet is
approaching at an angle of $30^\circ$--$70^\circ$ to the line-of
sight while the outer counter-jet is receding. Such an orientation
contradicts to the previously suggested models of the
inner jets and the bright arcs (e.g. Helfand et al.\ 2001).

We see thet the width of the outer jet, $\sim 3\times 10^{16}$ cm,
remains approximately the same along the jet in all observations.
This suggests an efficient confinement mechanism, perhaps
associated with magnetic fields generated by electric currents in
the pinched jet. The current required, $\sim 10^{32}\, e\, {\rm
s}^{-1}\sim 10^{12}\, {\rm amp}$, is an order of magnitude lower
than the Goldreich-Julian current in the pulsar magnetosphere. The
bright blobs and strong bends could be caused by the sausage and
kink instabilities, respectively, in such a pinched jet.

The spectrum of the outer jet fits well with a power law model
($\Gamma=1.3\pm0.1$) suggesting the synchrotron emission. The outer
 counter-jet also exhibits power-law spectrum with $\Gamma \approx 1.2$--1.5.
 Outside the bright PWN, there is an asymmetric, dim outer
diffuse nebula that is substantially brighter southwest of the
jet/counter-jet line. Its spectrum ($\Gamma \approx 1.5$) is softer than that of the
outer jet, but it is harder than the
spectrum of the brighter PWN shell ($\Gamma \approx 1.65$; region [6] in upper left panel of Fig.\ 1). Thus,
 it is possible that the X-ray
emitting particles in the dim
nebula are supplied through the outer jet.

This work was supported by SAO grant GO2-3091X.


\begin{figure*}
\centerline{\hbox{\psfig{figure=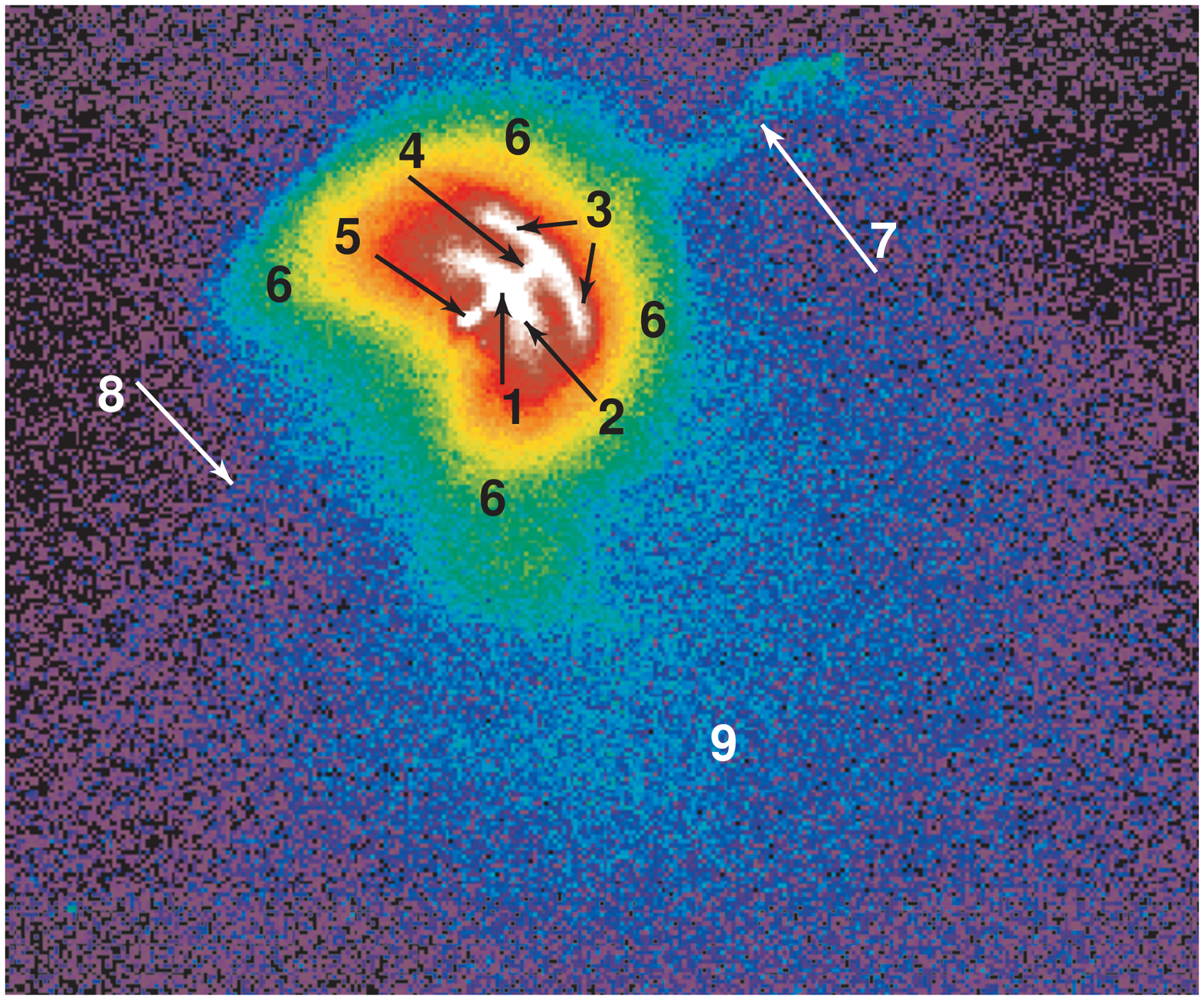,height=5.2cm,angle=0,clip=}\hspace{0.1cm}
\psfig{figure=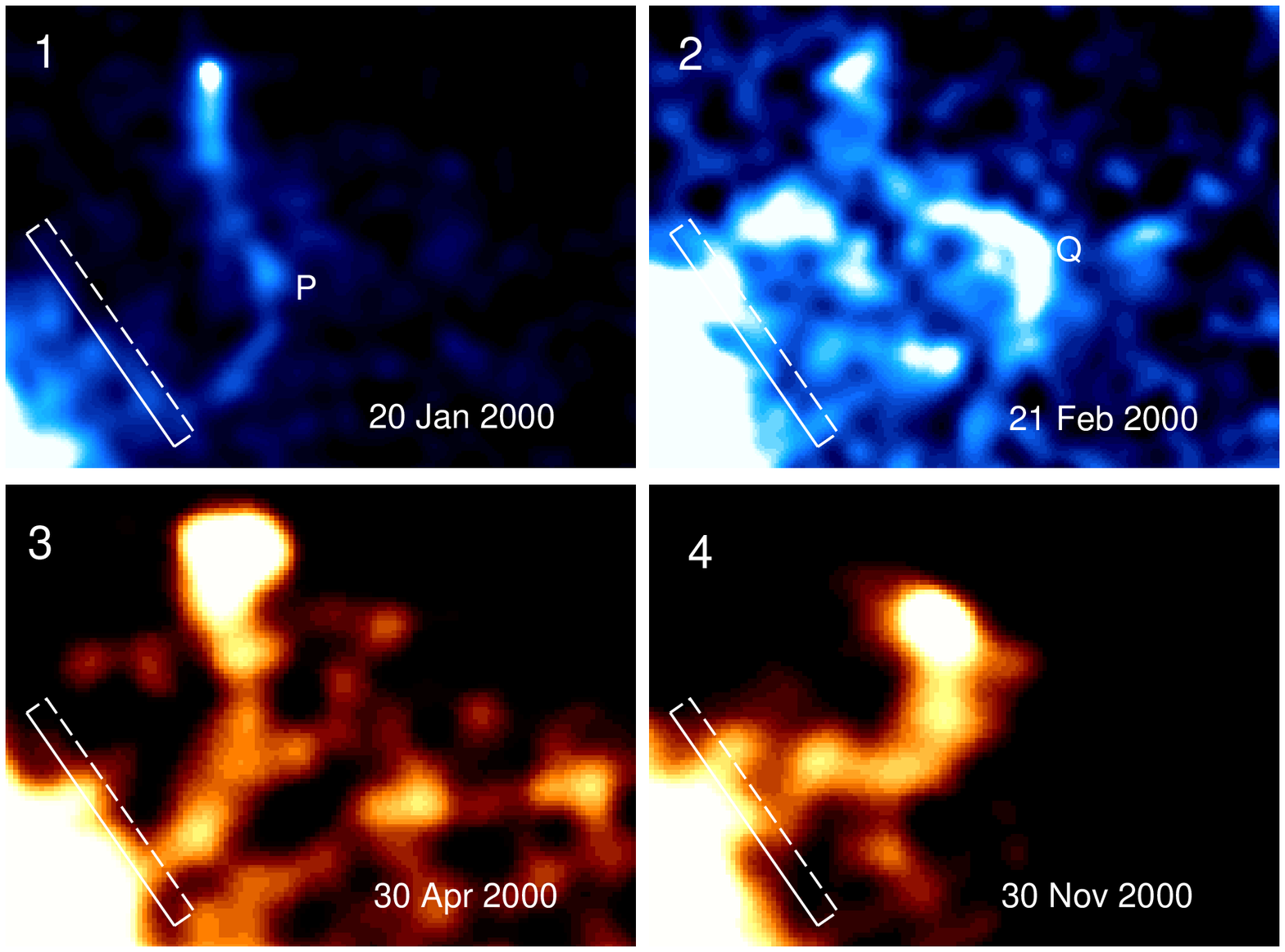,height=5.2cm,angle=0,clip=}}}\vspace{0.1cm}
\centerline{\hbox{\psfig{figure=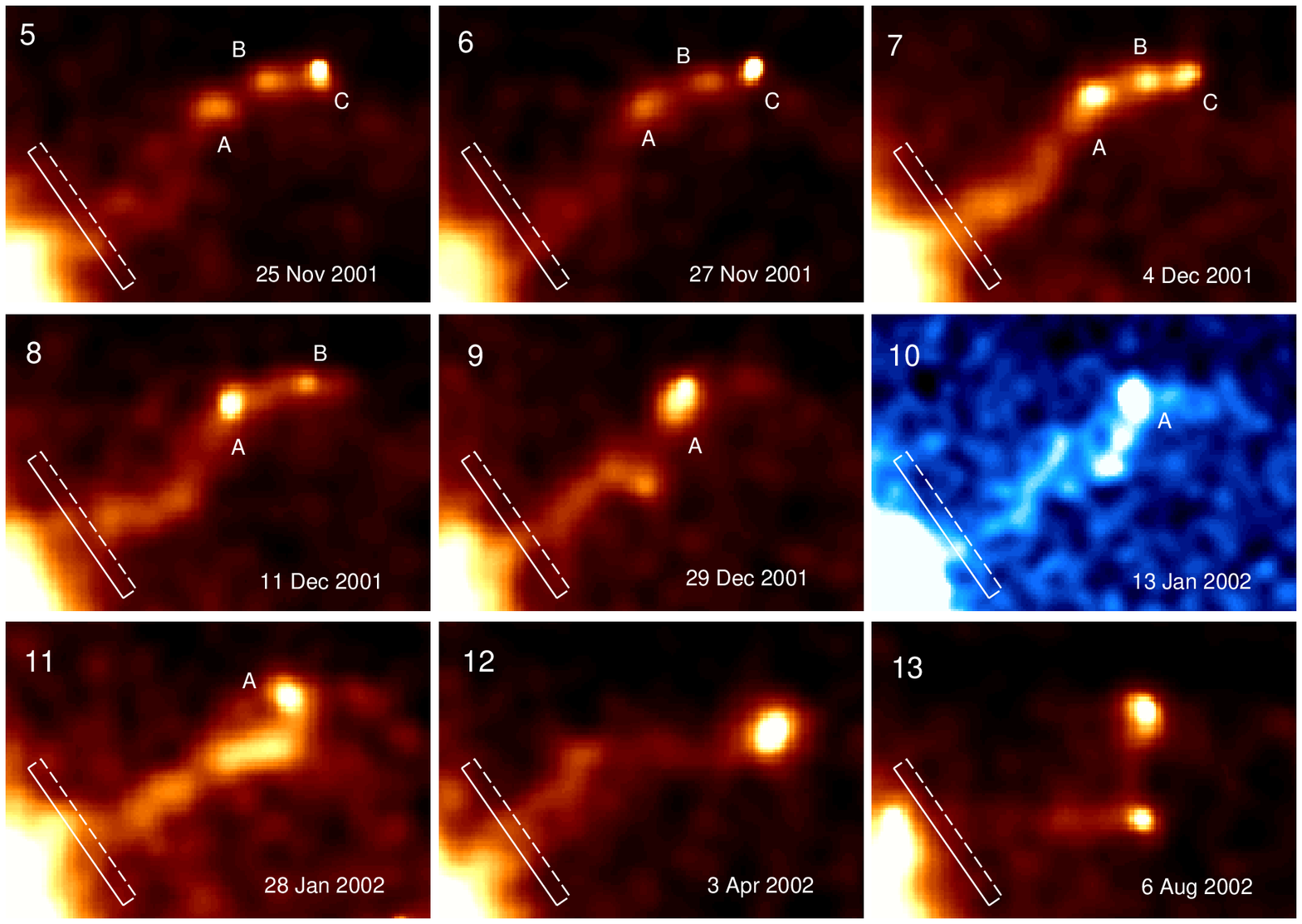,height=8.2cm,angle=0,clip=}}}
\caption{{\bf Top left:} The summed ACIS-S3 image of the
Vela PWN ($4\farcm 7\times 4'$) total exposure time about 160 ks).
{\bf Top right and bottom:} Sequence of \chan\ images of
the outer jet. Panels 1 2 and 10 are the HRC-I images; the rest are
 ACIS-S3 images.
 The size of each panels is $73''\times
53''$.
  The boxes, $28''\times2\farcs6$, at the same sky position
in all the panels, are overplotted to guide the eye.}
\end{figure*}

\end{document}